\documentclass[11pt]{article}

\usepackage[utf8]{inputenc}

\usepackage[a4paper,margin=1in]{geometry}
\usepackage{setspace}
\usepackage{amsmath,amssymb}

\usepackage{graphicx}
\usepackage{xcolor}
\usepackage{booktabs}
\usepackage{array}
\usepackage[font=small,labelfont=bf]{caption}

\usepackage{authblk}

\usepackage[round,authoryear]{natbib}
\usepackage[colorlinks=true,allcolors=blue!55!black,breaklinks=true]{hyperref}
\usepackage{url}
\title{\bfseries An Agentic Interface for End-to-End Probabilistic\\ Seismic Hazard and Risk Analysis}

\author[1]{Sreenath Vemula\thanks{Corresponding author: \href{mailto:vsreenath2@gmail.com}{\texttt{vsreenath2@gmail.com}}}}
\author[1]{Pierre Jehel}
\author[2,3]{Fabrice Cotton}
\author[1]{Filippo Gatti}

\affil[1]{Universit\'e Paris-Saclay, CentraleSup\'elec, ENS Paris-Saclay, CNRS, LMPS --- Laboratoire de M\'ecanique Paris-Saclay, 91190 Gif-sur-Yvette, France}
\affil[2]{GFZ Helmholtz Centre for Geosciences, Potsdam, Germany}
\affil[3]{University of Potsdam, Institute of Geosciences, Potsdam, Germany}

\date{}

\begin{document}
\maketitle

\begin{abstract}
\noindent Probabilistic seismic hazard and risk analyses are backbone to building codes, insurance pricing, and disaster management. Yet their open-engine pipelines remain accessible primarily to experts. We present the first agentic interface to the end-to-end probabilistic seismic hazard and risk chain via an open-source server, addressable through the Model Context Protocol (MCP). MCP wraps the OpenQuake engine and the 2020 European Seismic Hazard and Risk Models using twenty-four typed endpoints. Here, an agent is defined as a large language model (LLM) with tools. The LLM is confined to the role of an orchestrator so it plans, translates, and explains, while the OpenQuake engine and custom codes compute hazard values, damage probability, and loss. Each response carries source-model, ground-motion model (GMM), and certified data provenance for transparency. Results are benchmarked against ESHM20 at seventy-three cities; the replicated 475-year spectral accelerations match official values within a median of 5\,\%, and a full hazard-to-loss estimate runs in minutes. The interface additionally accepts a user-supplied empirical or machine-learning GMM on any tectonic region type of the published tree, and adds features that existing web services omit: conditional spectra, deterministic scenarios, surface hazard, per-building loss, retrofit comparison, and record selection with waveform retrieval. The proposed design layers transfer to other regional models, and to other hazards.
\end{abstract}

\medskip
\noindent\textbf{Significance.}\quad An LLM-powered chatbot accesses the full European probabilistic seismic hazard and risk models (ESHM20 and ESRM20). It reproduces the published reference at a median 5\,\% ln-unit level across multiple cities. Users can provide custom ground-motion models and run the regional logic tree with them in a single conversational session, chaining from hazard through to economic loss. The agent provides additional features beyond those the ESHM20/ESRM20 developers intended.


% ---------- Graphical abstract ----------
\begin{figure}[htbp]
  \centering
  \includegraphics[width=\linewidth]{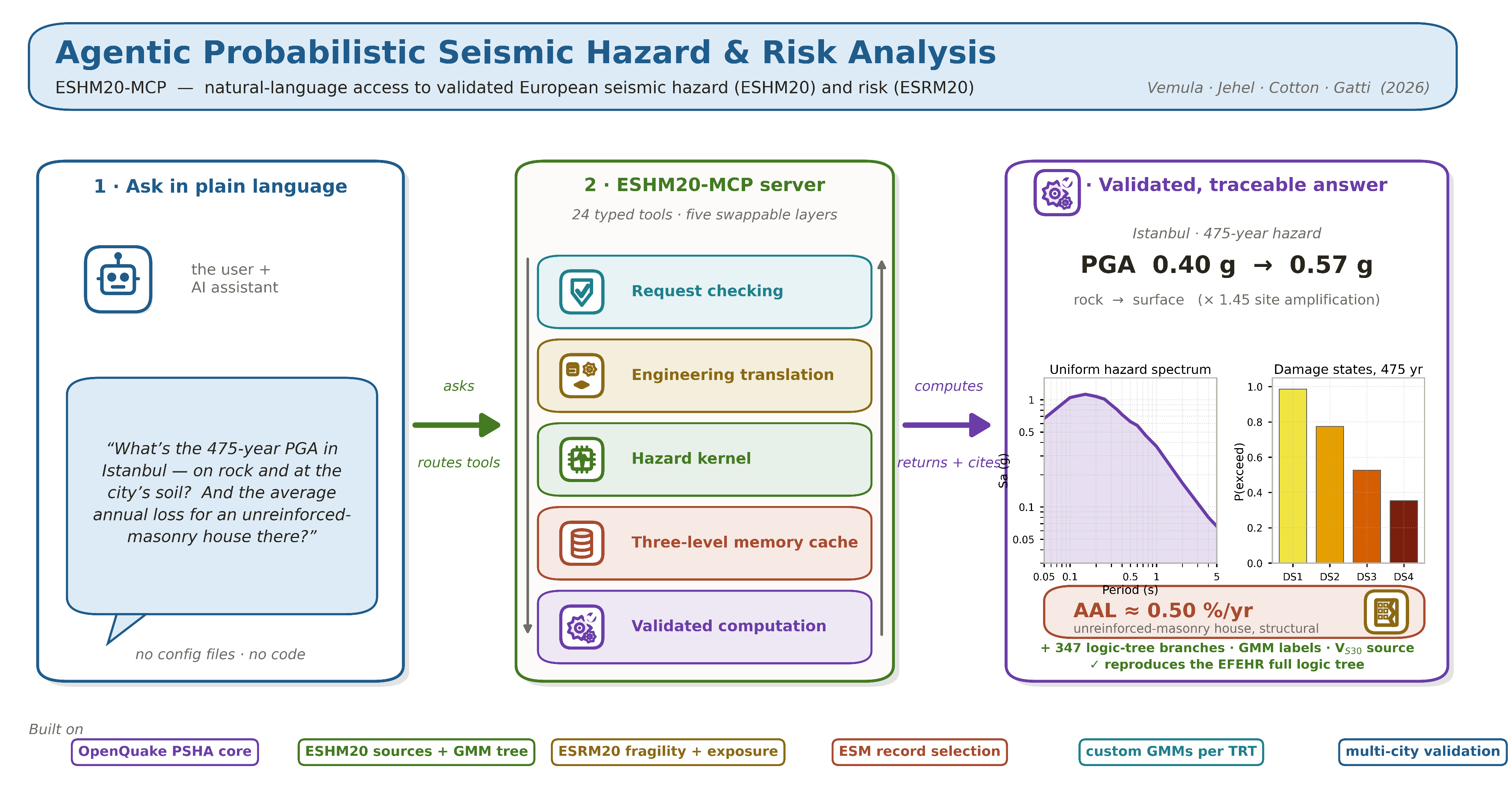}
  \caption*{\textbf{Graphical abstract.}}
\end{figure}

\section{Introduction}

Probabilistic seismic hazard and risk analyses are the basis for building codes, insurance underwriting, and emergency planning. Their open-data and open-source implementations have matured over the past few decades. Regional hazard models, exposure inventories, fragility libraries, and vulnerability functions now reach users through community repositories, and validated open engines such as OpenQuake \citep{pagani2014} run on these inputs at scale. End-to-end analysis nevertheless remains accessible only to specialists. To complete it, the practitioner must author Natural-hazards Risk Markup Language (NRML) configuration files, work through logic trees of tens to hundreds of weighted branches, map a building portfolio onto the Global Earthquake Model (GEM) taxonomy, supply or estimate a site shear-wave velocity \citep{heath2020}, and choose a conditioning period for the conditional mean spectrum \citep{baker2011}. While every step is well documented, an interface that ties them together within a single conversation with traceable provenance for each numerical output is missing. Additionally, the missing interface hinders non-specialists such as engineers, urban planners, and homeowners from using these models directly.

Updated European and regional GMMs continue to appear at a steady pace beyond the published regional logic tree \citep[e.g.,][]{manea2022,kotha2022,sreenath2023}. Evaluating how these new models affect the published hazard tree at hazard-map level is challenging. No packaged mechanism exists to substitute a candidate model into the regional logic-tree per tectonic type and then carry that substitution through hazard, fragility, and loss, and finally compare the results against previously published ones. Consequently, the gap between GMM development and its propagation to the engineering output (e.g., a PSHA or PSRA model) is also an interface gap.

A further interface barrier concerns how the obtained results are communicated. Hazard and risk outputs are read primarily as maps, and the choice of color palette and class boundaries measurably changes how a map is perceived. Color-vision-safe schemes are recommended over the still common rainbow palette \citep{schneider2023}. Existing services give the user very little control over this choice. An interface that can redraw the same hazard or risk product under different, perceptually sound palettes therefore improves comprehension as well as access.

Furthermore, large language models (LLMs) now draft research code, interpret experimental protocols, and answer technical questions across the natural and engineering sciences \citep{kim2025,pantiukhin2025}. However, without access to the NRML files, logic-tree weights, or integration routines, a question such as ``\emph{what is the average annual loss of a concrete frame in Athens?}'', when asked of an LLM, can be plausible yet unsupported \citep{lin2025,liY2025}. Agentic AI separates the responsibilities. Here, an agent is defined as the LLM in combination with the external tools, and is used as an orchestrator. The language model plans, translates, and explains, while every numerical step is delegated to a validated tool with schema-checked inputs and structured outputs \citep{gridach2025,hartung2025}; the LLM never originates a number.

The Model Context Protocol \citep{anthropic2024,hou2025} is the de facto standard for declaring tools to any compatible agent host, and it exposes specialised scientific capabilities across several fields. Among these are bioinformatics \citep{widjaja2025,flotho2026,xu2026}, hydrology \citep{yan2025,zhu2026}, and climate-data orchestration \citep{kim2025,kuznetsov2025}. Other examples span earth observation \citep{feng2026,chen2024}, structural design and health monitoring \citep{chenBao2025,geng2025,liang2025,liu2026}, and high-performance scientific computing \citep{pan2025,balis2026}. Engineering seismology lags this adoption. Only two systems have been reported, a ground-motion-field estimation agent \citep{zhao2026} and a pre-event disaster-simulation framework \citep{liL2025}, and neither covers the end-to-end probabilistic hazard and risk chain.

We close this interface gap with an end-to-end agentic server for probabilistic seismic hazard and risk analysis. The server is organised as five layers. A protocol adapter publishes the tools and checks every incoming request, while an engineering-domain translator turns practitioner vocabulary into model parameters. An engine-agnostic kernel dispatches the hazard computations, and a spatial cache reuses already-computed results at nearby sites. At the base, a validated numerical core performs every calculation; here OpenQuake is used. The architecture design works as a template. We demonstrate it on the 2020 European Seismic Hazard Model (ESHM20; \citealp{danciu2024}) and the companion 2020 European Seismic Risk Model (ESRM20; \citealp{crowley2021}).

The following are the key contributions of this work:
\begin{itemize}
  \item \textbf{Replication:} the server exposes the full hazard-to-loss chain through twenty-four typed endpoints in a single conversational session, so that any user can reproduce the published hazard maps, hazard curves, and spectra at any location and color scale. Every response also carries source-model and ground-motion provenance.
  \item \textbf{Substitution:} the user can replace individual links of the GMM chain, including a traditional or machine-learning GMM on any tectonic region type, and propagate the substitution to hazard curves, spectra, and maps against the published regional tree. This task previously required engine-level expertise.
  \item \textbf{Extensions} are provided beyond what ESHM20 and ESRM20 developers intended: conditional mean spectra, deterministic scenarios, surface (site-amplified) hazard, per-building damage and average annual loss, retrofit comparison, and conditional-spectrum record selection with waveform retrieval from the Engineering Strong-Motion (ESM) database \citep{mascandola2026}.
  \item Since the hazard and risk models carry significant uncertainty, this agentic framework can be used to perform sensitivity analysis at scale, though it is not explored in this work.
\end{itemize}

\section{Methods}

\subsection{Agentic AI and the MCP standard}

An agent in this work is defined as a language model wrapped in a control loop \citep{hartung2025}. It issues structured tool calls, reads the results, and revises its plan. Each tool calls a validated scientific code (e.g., OpenQuake) to run the computation, and each tool defines a JSON Schema for its inputs. Malformed or out-of-range inputs are therefore rejected before any computation runs. The Model Context Protocol (MCP) carries these messages in a fixed JSON-RPC 2.0 format, which lets any MCP-compliant client drive the server without writing custom code. The server is static, in the sense that the set of tools exposed to the language model does not change. It is also stateful: it remembers what the user queried earlier in the session and caches expensive results across calls.

\subsection{Server architecture}

\begin{figure}[htbp]
  \centering
  \includegraphics[width=\linewidth]{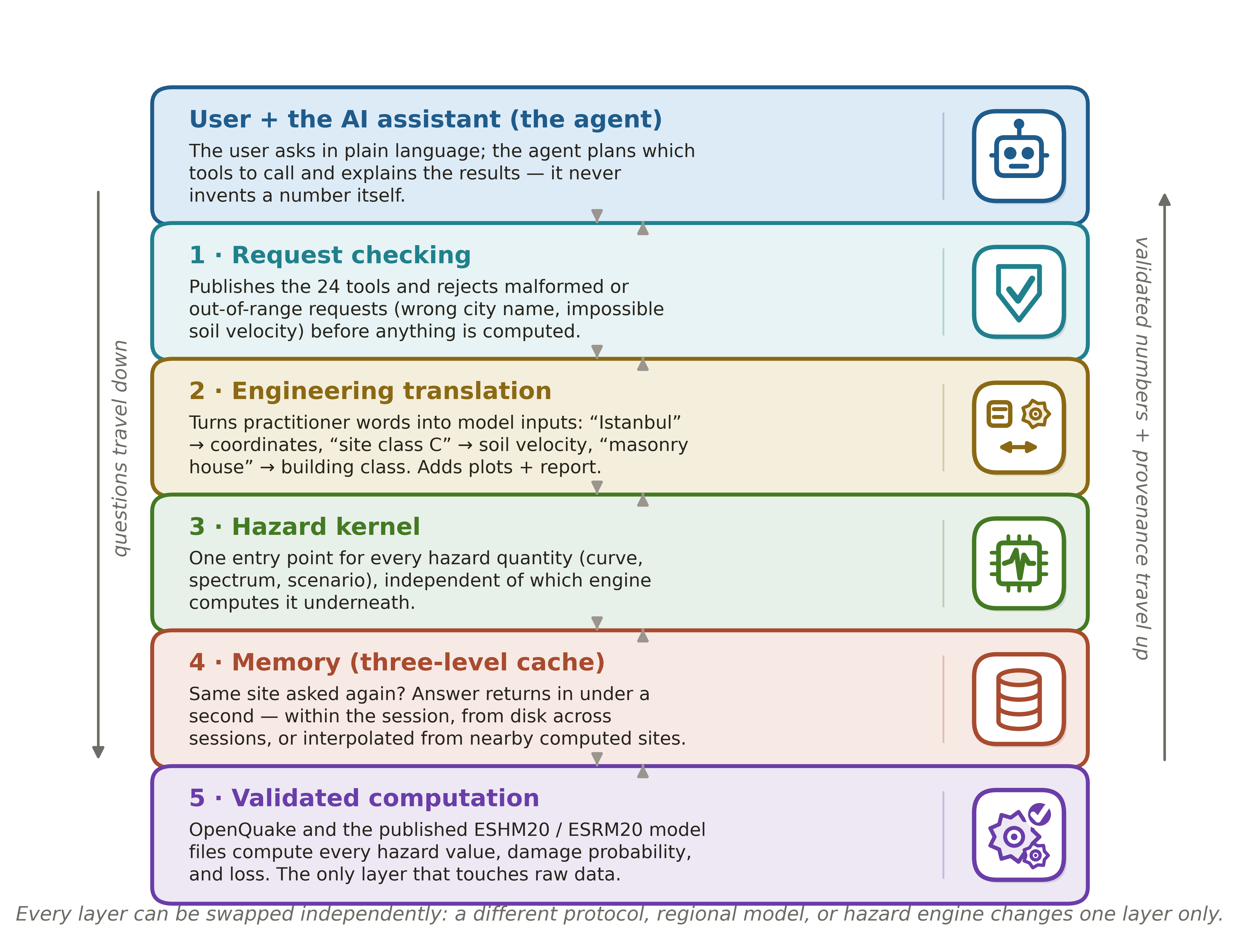}
  \caption{Layered organization of the server, from the conversation down to the validated numerical core. Each layer is described by what it contributes to a query rather than by its implementation: requests travel downward, validated numbers and their provenance travel back upward, and any layer can be substituted without affecting the others.}
  \label{fig:layers}
\end{figure}

As discussed, the design implements a Python process that speaks JSON-RPC 2.0 to any MCP-compliant client. The code is organized into five layers (Figure~\ref{fig:layers}), each of which carries a single responsibility and communicates only with its immediate layer neighbors. Replacing the engine, the regional code, or the protocol therefore affects only one layer at a time.

The outermost layer (L1) is a protocol adapter. L1 publishes the tool catalogue (e.g., \texttt{compute\_hazard}, \texttt{compute\_uhs}, \texttt{compute\_cms}, \texttt{estimate\_loss}) to the language model, which reads the catalogue to plan which tools to call. Every call the model emits is validated against its JSON Schema before any computation runs. Above it sits an engineering-domain translator (L2), which resolves a city to its coordinates using OpenStreetMap and determines the site shear-wave velocity ($V_{S30}$). An explicitly supplied $V_{S30}$ is used directly; otherwise an EC8 site class, if provided, is mapped to a representative $V_{S30}$ value; and if neither is given, $V_{S30}$ is queried automatically from the USGS Global $V_{S30}$ Mosaic \citep{heath2020} at the site coordinates. Additionally, colloquial building names are mapped to the corresponding GEM taxonomy, and a conditioning period appropriate to the structure is assigned. L2 also hosts the input--output layer: a session preferences store, a color-blind-safe plotting module, and an auto-accumulating site-grouped HTML report that every subsequent call inherits. Below the translator lies the engine-agnostic hazard kernel (L3), which exposes hazard curves, uniform hazard spectra, response spectra, conditional mean spectra, and single GMM evaluations.

A spatial cache (L4) is a software component that sits between the hazard kernel and the numerical core. It stores previously computed results so that later queries at the same or nearby sites can reuse them. The cache has two parts. An in-memory table holds the results of the current session, and once it reaches its size limit it discards its least recently used entries. The second part is an on-disk store, which keeps results across sessions, indexed by location rounded to a $0.1^{\circ}\times 0.1^{\circ}$ cell and by $V_{S30}$ rounded to a 50~m/s step. Each newly computed result is written to both parts, so the cache is updated on every new calculation. When later queries fall between stored cells, the cache returns a bilinear interpolation of the four nearest stored results. Repeated queries at a cached site are therefore returned almost instantly. Finally, L5 is the innermost layer, the numerical core. It reads the ESHM20 source-model files, which are in NRML format. Seismic hazard is computed in OpenQuake using the integral of \citet{cornell1968}. It also evaluates the ESRM20 fragility and vulnerability tables, fetches strong-motion records, and runs the record-selection optimization. The language model never interacts with the raw data directly.

\begin{figure}[htbp]
  \centering
  \includegraphics[width=\linewidth]{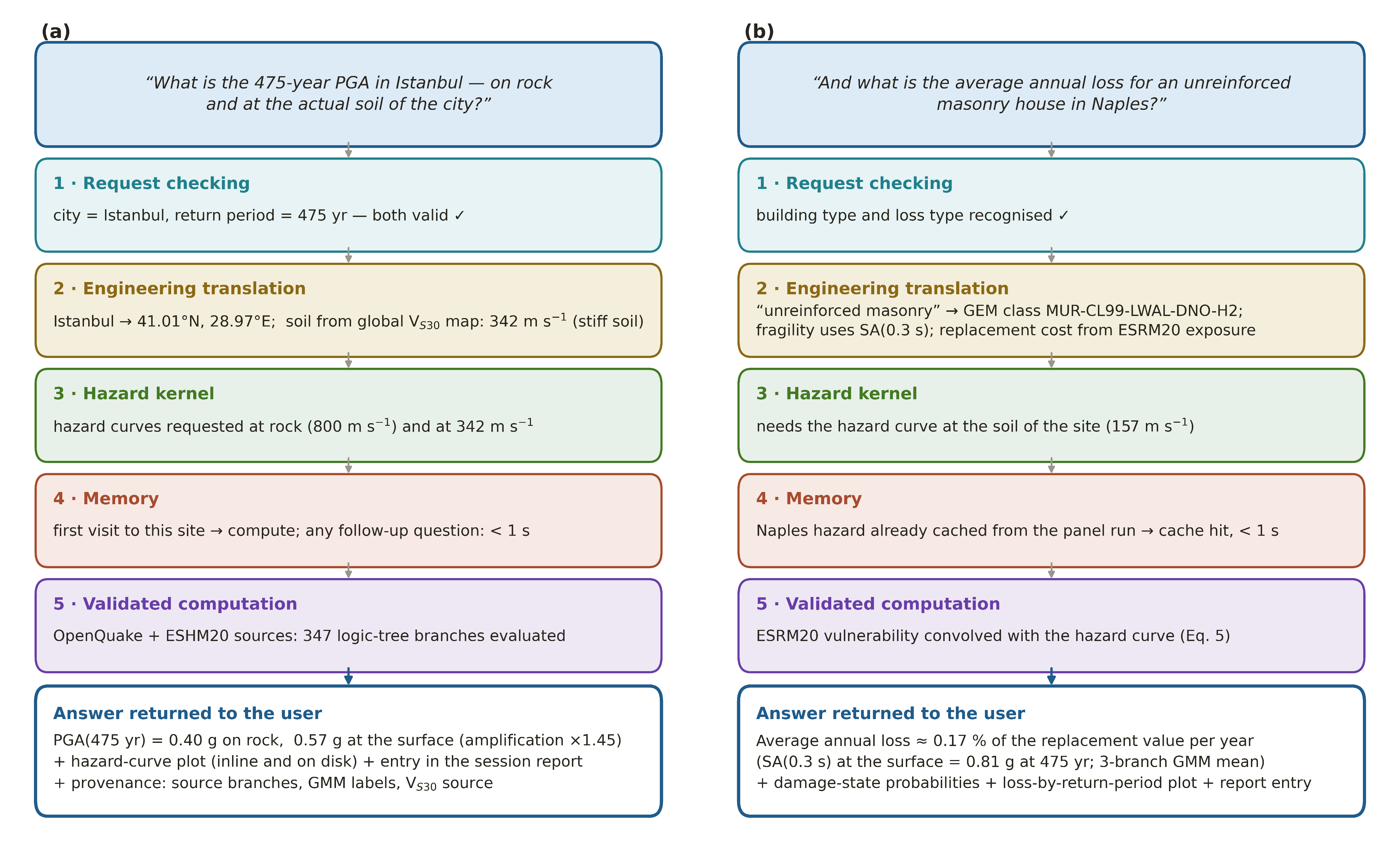}
  \caption{Two user journeys through the layered stack: (a) 475-year PGA at Istanbul, at rock and at the soil of the city; (b) average annual loss for an unreinforced-masonry dwelling at Naples, reusing its cached hazard from the panel run. Values are those returned in the verification session (Supplementary Table~S3).}
  \label{fig:journeys}
\end{figure}

Figure~\ref{fig:journeys} makes the division of labor concrete by tracing two queries through the layers, a hazard question at Istanbul and a loss question at Naples. Each box reports what the layer contributed in the demonstrated session, and the final boxes carry the values returned to the user; the loss query reuses the Naples hazard cached from the panel run, so it completes in under a second.

\begin{figure}[htbp]
  \centering
  \includegraphics[width=\linewidth]{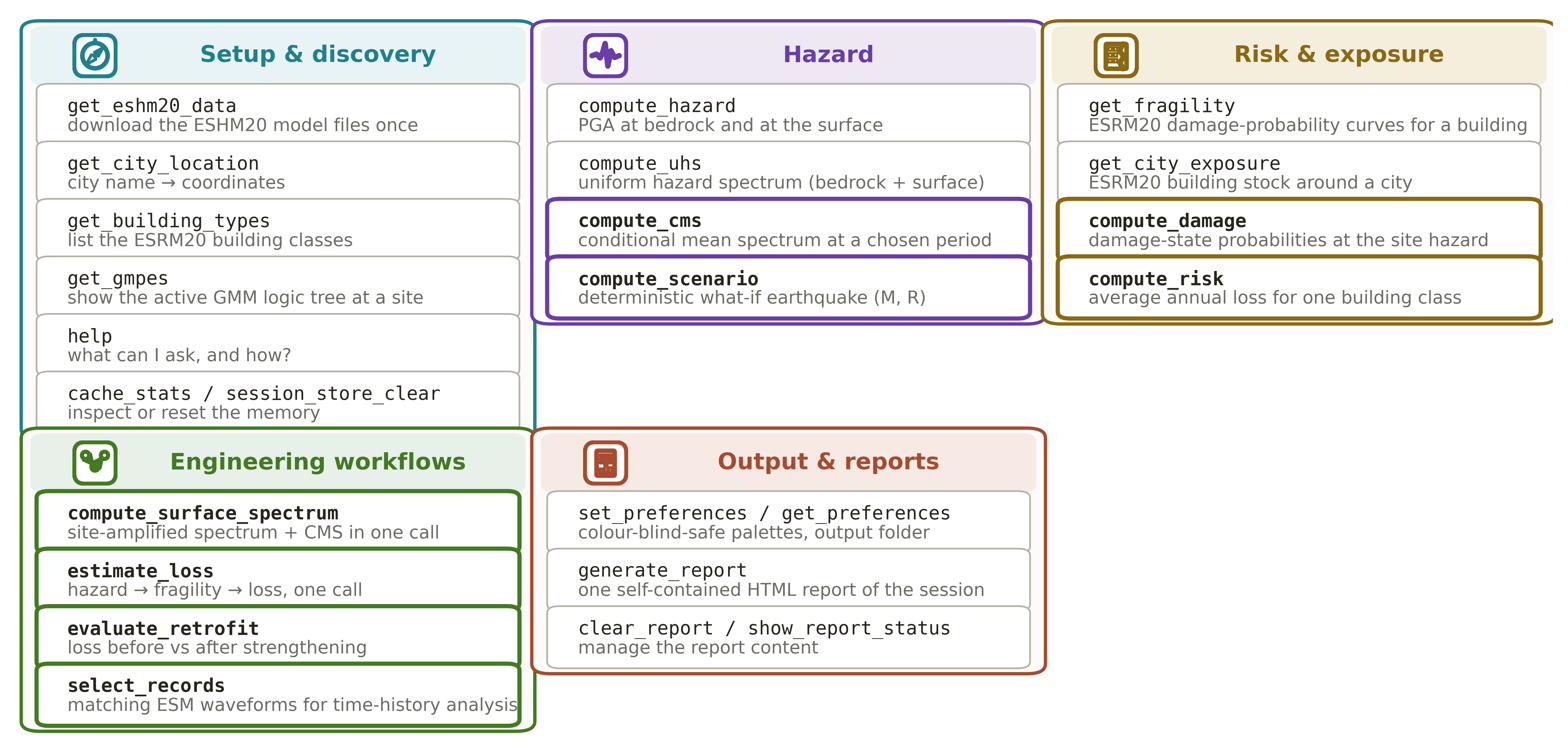}
  \caption{The twenty-four endpoints, grouped by user intent. \texttt{get\_} endpoints return published ESHM20/ESRM20 content; \texttt{compute\_} endpoints run new calculations; bold boxes mark capabilities not available from the EFEHR web services. Every hazard endpoint also accepts a per-tectonic-region GMM override.}
  \label{fig:endpoints}
\end{figure}

The agent sees a catalogue of twenty-four typed endpoints organized by user intent (Figure~\ref{fig:endpoints}). Every endpoint is published by the protocol adapter (L1), which validates the incoming call. The call then descends through the engineering-domain translator (L2), the hazard kernel (L3), the spatial cache (L4), and the numerical core (L5), and the validated result returns along the same path. Names follow a fixed convention. Endpoints prefixed \texttt{get\_} imply that they simply fetch published content from ESHM20 or ESRM20: model files, taxonomies, fragility tables, exposure, and the active GMM logic tree. Endpoints prefixed \texttt{compute\_}, on the other hand, imply that a calculation is performed to obtain results. Setup and discovery endpoints prepare the session and expose the catalogue and active settings. Hazard-group endpoints traverse the probabilistic chain from uniform hazard spectra (bedrock and surface) to the conditional mean spectrum \citep{baker2011} and a deterministic scenario, while risk and exposure endpoints turn intensity into consequence through fragility, damage-state probabilities, the loss-ratio integral, and exposure. Four engineering workflows combine these endpoints into the questions a practitioner actually asks (surface spectrum with optional conditional mean spectrum, loss estimation, retrofit evaluation, and ESM record selection using \citet{bakerLee2018} and \citet{jayaram2011}), and a fifth group manages preferences and the HTML report.

It must be reminded that the hazard curves, spectra, and maps (only at the bedrock level) are provided by European Facility for Earthquake Hazard and Risk (EFEHR) web services \citep{danciu2024}. The remaining endpoints, in bold in Figure~\ref{fig:endpoints}, are not available through those services. A small number of tools exposed to the language model significantly reduces the risk of hallucination and wrong tool call.

\subsection{Data substrate}

ESHM20 contains the area-source, fault-source, and smoothed-seismicity model files in NRML, paired with several GMM choices. For the Vrancea intermediate-depth seismicity it adopts the BC-Hydro model \citep{abrahamson2016}. The shallow-crustal regime uses the \citet{kotha2020,kotha2022} ESHM20 backbone, whereas the Cratonic regime relies on the \citet{weatherillCotton2020} regionally adapted scaled-backbone tree. Volcanic source zones use the \citet{lanzanoLuzi2019} GMM. ESRM20 supplies the lognormal fragility tables, the mean loss-ratio functions, and the capacity-curve library \citep{martinsSilva2020}. Two further datasets complete the set: a continental exposure database harmonized against the SERA D26.3 deliverable \citep{crowley2020,dabbeek2021}, and the regional site-amplification model \citep{weatherill2023}. Ground-motion records used by the record-selection workflow are obtained from the ESM database. All probabilistic seismic hazard analysis runs in OpenQuake hazardlib.

\subsection{Custom GMM per tectonic region}

By default, every hazard call uses the published ESHM20 GMM logic tree. A single optional argument replaces the model for any of the seven tectonic regions (shallow default, craton, subduction interface, subduction inslab, non-subduction deep, volcanic, Iceland Atlantic active). The replacement can be an OpenQuake-registered model, a user-supplied empirical or machine-learning GMM function, or a regional logic-tree file. Regions left untouched keep their ESHM20 defaults, and the active tree with branch weights is logged on every call, so results from a candidate model can be compared directly against the published tree. The input contract is given in Supplementary~S5.

\subsection{Output products and session reporting}

Every numerical endpoint returns three items. One is a JSON file holding the numerical results together with provenance metadata: source-model branches, GMM labels, and data sources. Another is a Matplotlib figure, emitted as a base-64 PNG inline in the response, with a copy also written to disk under a relative session folder. The last is an entry in the session report buffer. A session-wide preferences endpoint sets this behavior once, and every subsequent call inherits it.

The plotting layer offers five palettes through a color-mode argument. These include the \citet{wong2011} color-blind-safe default, along with modes for deuteranopia, protanopia, tritanopia, and monochrome printing, each perceptually uniform under the corresponding color-vision deficiency. Visualization research on seismic hazard maps shows that palette and classification scheme measurably change how hazard is read, and it recommends perceptually ordered, color-vision-safe schemes \citep{schneider2023}. A session-wide report endpoint gathers every figure, value table, and one-line interpretation. The result is a single self-contained HTML document, grouped by site, that can be archived alongside the underlying JSON.

\subsection{Governing equations}

The mean annual rate of exceedance $\lambda(s)$ summarizes seismic hazard at the site for a chosen intensity measure (e.g., peak ground acceleration or spectral acceleration) at a structural period \citep{cornell1968}. The intensity at a target return period $T_R$ follows from log-linear interpolation of the hazard curve between bracketing levels $s_j$ and $s_{j+1}$ with annual rates $\lambda_j$ and $\lambda_{j+1}$ (Eq.~\ref{eq:hazard}):
\begin{equation}
\begin{gathered}
\lambda(s) = -\ln\!\bigl(1 - P(S > s \mid 1~\text{yr})\bigr), \qquad \lambda^{*} = \frac{1}{T_R},\\[4pt]
s(T_R) = s_j\left(\frac{s_{j+1}}{s_j}\right)^{\!f}, \qquad
f = \frac{\ln(\lambda_j/\lambda^{*})}{\ln(\lambda_j/\lambda_{j+1})}.
\end{gathered}
\label{eq:hazard}
\end{equation}
Spectral displacement follows from $S_D(T) = g\,T^{2}/(4\pi^{2})\,S_A(T)$. Applying Eq.~\ref{eq:hazard} at the EC8 spectral periods yields the uniform hazard spectrum.

The uniform hazard spectrum is obtained by reading the hazard curve at the same return period independently at each structural period. It therefore assumes that the target intensity is reached at every period at the same time, which is unrealistic \citep{baker2011}. The conditional mean spectrum is a more realistic target which fixes the intensity at one conditioning period $T^{*}$, typically the fundamental period of the building under analysis, and then gives the response spectrum expected to occur together with that intensity at every other period $T_i$. Because spectral values at different periods are correlated, the expected value at any other period $T_i$ falls below the uniform hazard value by an amount set by the inter-period correlation $\rho(T_i, T^{*})$ of \citet{bakerJayaram2008}. Equation~\ref{eq:cms} gives this expected spectrum (the conditional mean) and Equation~\ref{eq:cmsvar} gives its variance:
\begin{equation}
\ln \mathrm{CMS}(T_i) = \mu_{\ln}(T_i \mid M, R) + \rho(T_i, T^{*})\,\varepsilon^{*}\,\sigma_{\ln}(T_i),
\label{eq:cms}
\end{equation}
\begin{equation}
\sigma_{\ln}^{2}(T_i \mid T^{*}) = \sigma_{\ln}^{2}(T_i)\bigl(1 - \rho^{2}(T_i, T^{*})\bigr).
\label{eq:cmsvar}
\end{equation}
In Equations~\ref{eq:cms} and~\ref{eq:cmsvar}, $\mu_{\ln}(T_i \mid M, R)$ and $\sigma_{\ln}(T_i)$ are the mean and standard deviation of the natural-log spectral acceleration predicted by the ground-motion model at period $T_i$ for the controlling magnitude $M$ and distance $R$, $\varepsilon^{*}$ is the number of standard deviations by which the target intensity exceeds the median at the conditioning period $T^{*}$, and $\rho(T_i, T^{*})$ is the correlation between the log spectral accelerations at periods $T_i$ and $T^{*}$.

Damage-state exceedance probabilities follow a lognormal cumulative distribution in the intensity measure (Eq.~\ref{eq:fragility}), with median $\theta_k$ and dispersion $\beta$ tabulated in the ESRM20 fragility library:
\begin{equation}
P(DS \geq ds_k \mid S = s) = \Phi\!\left(\frac{\ln s - \ln \theta_k}{\beta}\right).
\label{eq:fragility}
\end{equation}
Differencing the cumulative form gives discrete damage-state probabilities. Convolving the mean loss-ratio function $L(s)$ with the hazard-curve slope at the building's intensity-measure period yields the average annual loss as a fraction of the replacement cost (Eq.~\ref{eq:aal}; \citealp{martinsSilva2020}), with trapezoidal quadrature integrating the full hazard-curve sampling:
\begin{equation}
\mathrm{AAL} = \int_{0}^{\infty} L(s)\,\left|\frac{d\lambda(s)}{ds}\right| ds.
\label{eq:aal}
\end{equation}
Applying Eq.~\ref{eq:aal} to an as-is and a retrofitted version of the same building gives the annual monetary benefit of the retrofit, $B = (\mathrm{AAL}_{\text{as-is}} - \mathrm{AAL}_{\text{retrofit}})\,C_{\text{repl}}$; the replacement cost $C_{\text{repl}}$ comes from the ESRM20 exposure tables or a user override. Default parameter values for logic-tree branch counts, cache thresholds, the return period, and damping are listed in Table~S1.

\subsection{Interfaces}

Two interfaces share the same session state. The first is a JSON-RPC entry point for any MCP-compatible host (e.g., Claude Code, Codex, VS Code). A locally run browser interface provides the second, combining a form-based workspace, a chat panel, and a one-click HTML report. A study begun in one interface continues in the other without state loss. Representative prompts, with the endpoints each one invokes, are collected in Supplementary Table~S6.

\section{Results}

\subsection{Verification overview}

Initially, implementation correctness is verified, testing each governing equation against an external reference. Eq.~\ref{eq:hazard} is verified against analytic exceedance solutions, Eqs.~\ref{eq:cms} and~\ref{eq:cmsvar} against the tabulated correlation coefficients of \citet{bakerJayaram2008}, Eq.~\ref{eq:fragility} against the ESRM20 spreadsheet libraries, and Eq.~\ref{eq:aal} against a closed-form integral oracle, with the numerical agreement summarized in Table~S2. Hazard accuracy is verified by comparing the server output with the ESHM20 pan-European reference at a multi-city European panel and two engineering return periods. This comparison runs using the published ESHM20 GMM logic tree.

\subsection{Hazard distribution across Europe}

\begin{figure}[htbp]
  \centering
  \includegraphics[width=\linewidth]{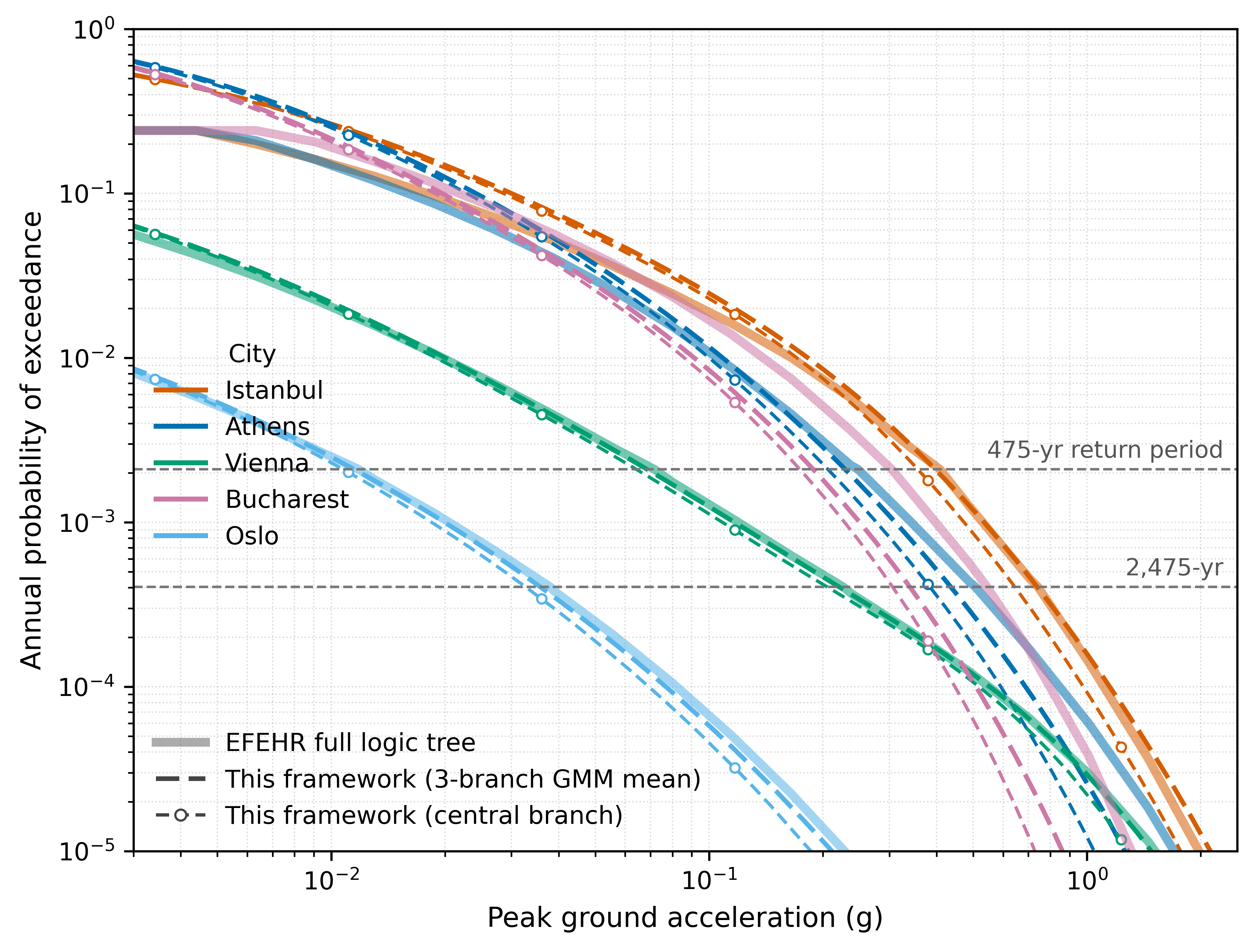}
  \caption{PGA hazard curves at representative cities, $V_{S30}=800$~m\,s$^{-1}$: this framework as a three-branch Gauss--Hermite ground-motion mean (dashed) and central branch (dashed with markers) against the EFEHR service evaluating the complete ESHM20 tree (solid). Agreement is within ${\sim}8\,\%$ at the 475-year level for the three-branch mean except at Vrancea-dominated Bucharest, where even the three-branch mean omits influential deep-seismicity branches.}
  \label{fig:hazcurves}
\end{figure}

\begin{figure}[htbp]
  \centering
  \includegraphics[width=\linewidth]{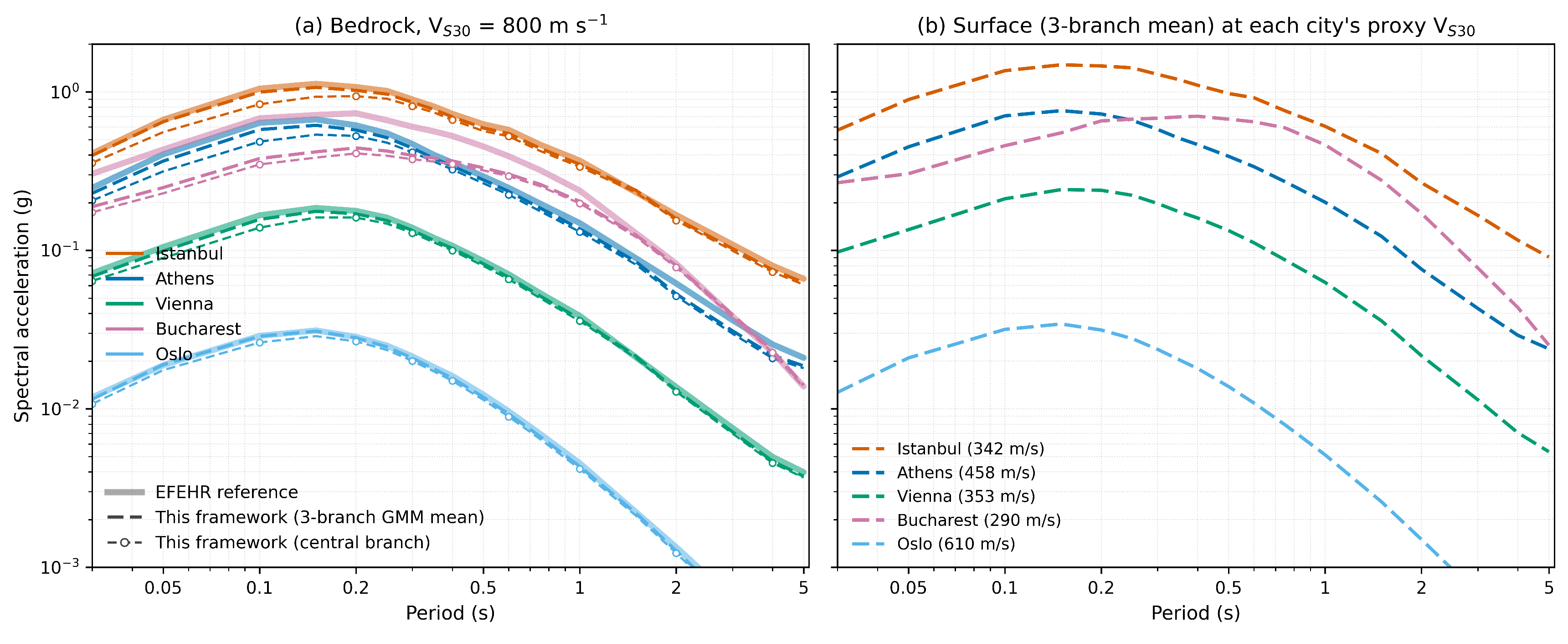}
  \caption{475-year uniform hazard spectra at the panel cities. (a) Rock reference: this framework as a three-branch Gauss--Hermite ground-motion mean (dashed) and central branch (dashed with markers) against the EFEHR full-tree reference (solid). (b) Surface spectra at each city's proxy-derived $V_{S30}$, a site-amplified product the EFEHR services do not provide.}
  \label{fig:uhs}
\end{figure}

We first reproduce the peak ground acceleration map at the 475-year return period (Supplementary Figure~S1), and the same endpoints return the underlying hazard curves and spectra at any chosen site (Figures~\ref{fig:hazcurves} and~\ref{fig:uhs}). The spatial pattern follows the regional structure of the ESHM20 source model. Hazard peaks along the Hellenic and Calabrian arcs, the Apennines, eastern Anatolia, the Vrancea zone, and the Icelandic rift (PGA $>0.30$~g), then falls to the stable north and west of the continent ($<0.05$~g).

Hazard curves are a sharper replication test, since they expose the full rate-intensity relationship at a site. Figure~\ref{fig:hazcurves} compares the PGA hazard curves computed by the server at representative panel cities (Istanbul, Athens, Vienna, Bucharest, and Oslo) at the rock reference $V_{S30}=800$~m/s. The reference set uses the EFEHR hazard service, which evaluates the complete ESHM20 logic tree using 10{,}000 Monte-Carlo runs. Across the shallow-crustal and cratonic sites the server and full-tree curves nearly coincide over the engineering range. At the 475-year return period the server PGA lies within 3\,\% of the full-tree value at Oslo (0.011 versus 0.012~g), 4\,\% at Vienna (0.068 versus 0.072~g), 3\,\% at Istanbul (0.40 versus 0.41~g), and 8\,\% at Athens (0.23 versus 0.25~g). Bucharest behaves differently. There, even the three-branch ground-motion mean under-predicts the full tree by 38\,\% (0.19 versus 0.30~g at 475 years).

The same replication extends to spectra. Figure~\ref{fig:uhs}a shows the 475-year uniform hazard spectra at the same cities on the rock reference, with the EFEHR full-tree spectra overlaid once more. The agreement follows the hazard curves, and the Bucharest spectrum carries the Vrancea under-prediction across periods. Figure~\ref{fig:uhs}b gives the corresponding spectra at the surface, computed with each city's proxy-derived shear-wave velocity, a product the existing web services do not provide. Site amplification raises the peak-ground-acceleration ordinate by factors of about 1.1 (Oslo, 610~m\,s$^{-1}$) to 1.45 (Istanbul, 342~m\,s$^{-1}$), and it is this surface spectrum that the risk endpoints consume. At Oslo, the visible long-period divergence sits below 0.01~g, where the three-branch cratonic mean and the tail of the reference grid dominate a log-scale comparison without engineering consequence.

\subsection{Hazard accuracy}

We compared the 5\,\%-damped acceleration response spectra produced by the server against the ESHM20 pan-European reference at seventy-three European cities. These cities span the tectonic envelope of the continent: the stable Cratonic interior, the Atlantic margin, active shallow crust, Aegean and Mediterranean subduction, Vrancea intermediate-depth seismicity, and the Icelandic volcanic margin. Both comparisons were made at the 475-year and 2{,}475-year return periods. Discrepancy is quantified by the mean absolute error in natural-log spectral acceleration, taken over peak ground acceleration and the sixteen periods between 0.05~s and 5~s common to both spectra. Whereas the EFEHR reference evaluates the complete ESHM20 logic tree, the server evaluates the three-branch Gauss--Hermite ground-motion mean. The residual therefore aggregates the higher-order ground-motion branches together with any implementation difference \citep[comparison strategy after][]{weatherill2024,abbott2020}. Across the cities, the replicated spectra agree with the official values within a median absolute ln-residual of 0.05 at the 475-year return period (about 5\,\%; mean 0.07) and 0.06 at the 2{,}475-year return period (about 6\,\%; mean 0.08). For 97\,\% of cities the residual stays below 0.25 ln-units (about 28\,\%; Supplementary Figure~S2).

The spatial pattern of the residual is physically interpretable. Most European sites fall within the shallow-crustal branch of the ESHM20 logic tree, where the default configuration captures the dominant epistemic uncertainty of the tree \citep{danciu2024}. The two largest residuals occur at Chi\c{s}in\u{a}u (0.48) and Bucharest (0.31), both Vrancea sites whose long-return-period hazard is dominated by intermediate-depth (60 to 180~km) earthquakes routed through the BC-Hydro non-subduction-deep GMM. Because the default subset evaluates the deep source-model logic tree in full, every maximum-magnitude branch is loaded at its aggregated logic-tree weight. The ground-motion logic tree, however, is represented only by its three-branch Gauss--Hermite mean. Consequently, the residual against ESHM20 reflects the higher-order GMM epistemic branches that the three-point quadrature does not capture, and it captures them precisely at the sites where those branches matter most.

The next-largest residual occurs at low-hazard cratonic Berlin (0.23), which sits at the opposite tectonic extreme. Its absolute spectral acceleration at 1~s falls below 0.005~g at the 475-year return period, well below any code-relevant threshold, so small absolute differences inflate the natural-log metric out of proportion to their engineering significance. Dublin and Madrid, judged the same way, now agree to within 0.08 and 0.06 ln-units. Scatter rather than systematic bias dominates the bulk residual: the mean signed ln-residual is only $-0.04$ (a 4\,\% net under-prediction) against a mean absolute residual of 0.07, so the per-city errors are near-symmetric and largely cancel in aggregate. By default the configuration evaluates only the highest-weighted source and ground-motion branches. Sampling the full ESHM20 logic tree more densely, for example by Monte-Carlo over its area-source, fault-source, and ground-motion branches, would converge toward the official EFEHR full-tree values. Because the three-branch ground-motion mean already captures most of this epistemic spread, the residual reflects the omitted higher-order branches rather than an implementation deficit. The small net under-prediction has a simple origin. A full ground-motion logic-tree mean lies above its median branch by the lognormal factor $\exp(\sigma_\mu^{2}/2)$ for an epistemic spread $\sigma_\mu$, and averaging the three highest-weighted branches recovers most, but not quite all, of that upward shift. A direct test isolates the cause. Switching from the single central branch to the three-branch Gauss--Hermite mean lowers the median 475-year error from 9.3\,\% to 4.7\,\% and improves 82\,\% of cities, so the ground-motion quadrature controls the residual. The optional five-point mode restores the two outer Gauss--Hermite nodes (at $\pm 2.857\,\sigma_\mu$) that the three-point mean omits, which most reduces the Vrancea and Craton residuals.

\subsection{Incorporating new models into the hazard calculation chain}

To show that an externally developed model can be substituted into the published tree and propagated end-to-end, we re-ran two representative high-hazard sites through the custom-GMM interface. At Istanbul, the shallow-crustal regime routes through a custom interpretable machine-learning GMM \citep{sreenath2025}, which replaces the \citet{kotha2020,kotha2022} backbone. Bucharest is governed by the non-subduction-deep regime, which carries the Vrancea seismicity that controls Romanian and Moldovan hazard. There the calculation routes instead through the regional model of \citet{manea2022}, replacing the BC-Hydro slab variant. Every other tectonic region retains its ESHM20 defaults.

\begin{figure}[htbp]
  \centering
  \includegraphics[width=\linewidth]{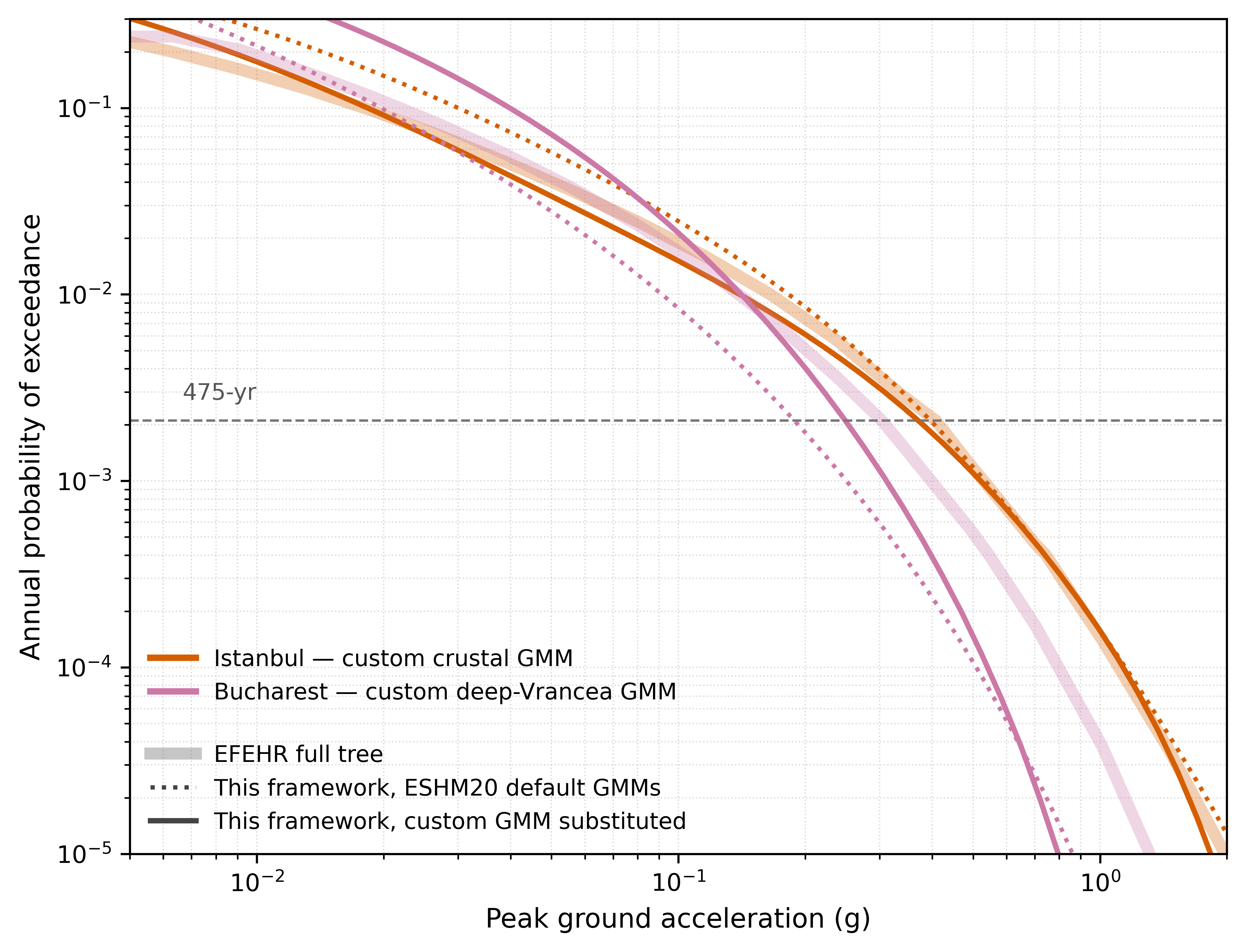}
  \caption{PGA hazard curves under two custom-GMM substitutions ($V_{S30}=800$~m\,s$^{-1}$). Istanbul: the shallow-crustal regime routed through a custom interpretable machine-learning GMM \citep{sreenath2025}. Bucharest: the Vrancea intermediate-depth regime routed through the regional \citet{manea2022} model. Solid: custom GMM substituted; dotted: default ESHM20 GMMs; wide translucent: EFEHR full tree; dashed: 475-year return period. Each substitution is propagated through the full PSHA while all other tectonic regions keep their ESHM20 defaults. The EFEHR full tree evaluates the default BC-Hydro and Kotha models and is shown only as the published-tree reference.}
  \label{fig:customgmm}
\end{figure}

Figure~\ref{fig:customgmm} shows the effect directly on the PGA hazard curves at the two demonstration sites. At Bucharest, substituting the regional \citet{manea2022} model for the BC-Hydro slab variant raises the 475-year hazard from 0.19~g to 0.25~g, the difference between two published intermediate-depth models propagated through the full PSHA by a single argument. The EFEHR full tree still evaluates BC-Hydro, not Manea, so it is shown only as the published-tree reference, not as a target the substitution is meant to match.

At Istanbul the \citet{sreenath2025} GMM returns a 475-year PGA of 0.37~g, against 0.40~g for the default Kotha backbone. Although these two shallow-crustal models were developed independently, they agree to within about 10\,\%. The point is not whether either matches the EFEHR reference, but that an externally trained model can be dropped into the published tree and carried through the full hazard calculation in a single call.

A developer can swap a candidate model into the published tree and then read off the resulting hazard curve, spectrum, or map, all within the same conversational session that produced the default-tree results.

\subsection{Adding new features: conditional spectra and record selection}

\begin{figure}[htbp]
  \centering
  \includegraphics[width=\linewidth]{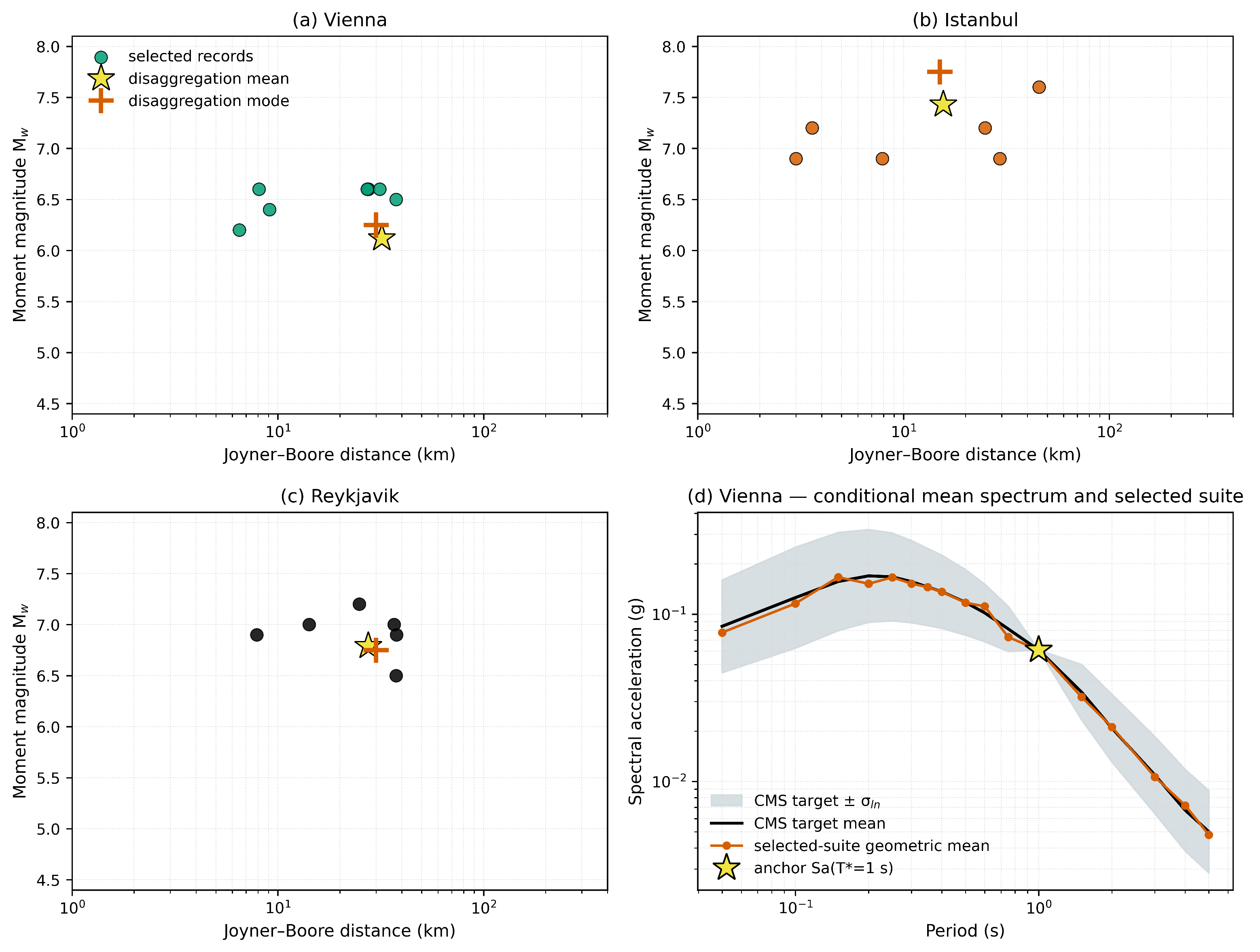}
  \caption{Disaggregation, conditional spectrum, and record selection. (a--c) Selected records (dots) against the disaggregation mean (star) and mode (cross) at Vienna, Istanbul, and Reykjav\'ik, with the spectral-shape error norm $\varepsilon$ in the titles. (d) Conditional mean spectrum at Vienna ($\pm\sigma$ band) with the selected-suite geometric mean; neither product is available from the EFEHR services.}
  \label{fig:recordsel}
\end{figure}

We ran the record-selection workflow at the capital-city panel, drawing seven records per site under disaggregation-anchored default filters. At the six low-to-moderate-hazard capitals, the defaults succeeded on the first attempt. Three shallow-crustal sites have sparser database coverage, which triggered automatic loosening of the bounds; Supplementary Table~S4 tabulates the per-city stages. The spectral-shape error norms in Figure~\ref{fig:recordsel} track the database. Because shallow-crustal sites at moderate distance are well represented in the ESM holdings, they achieve the lowest norms. The Vrancea-dominated Bucharest target reaches the highest norm, since its deep-focus, far-field scenarios (magnitude $\approx 7.6$, Joyner--Boore distance $\approx 190$~km) are sparse in the database. At such sites the workflow reports the loosening stage and candidate-pool count plainly, so the practitioner can decide whether to broaden the filters or escalate to a region-specific simulation suite.

Figure~\ref{fig:recordsel} condenses the workflow into one view: panels (a--c) compare the selected records with the disaggregation target at Vienna, Istanbul, and Reykjav\'ik, and panel (d) overlays the target conditional mean spectrum (Eqs.~\ref{eq:cms}--\ref{eq:cmsvar}), a product the existing services do not provide, with the geometric mean of the suite selected at Vienna. There the geometric mean lies within the conditional $\pm\sigma_{\ln}$ band at every period. The Vrancea-dominated Bucharest target, by contrast, falls outside the design population of the database: deep-focus, far-field scenarios are structurally sparse, and the selected-suite spectral shape therefore cannot match the conditional mean at short periods regardless of which conditional-spectrum selector is applied; the high error norm in Table~S4 quantifies this gap.

\subsection{End-to-end demonstration: computing risk}

An end-to-end loss-estimation run at the capital-city panel confirms that the modules behave consistently when chained. Each city receives a taxonomy representative of its residential stock: mid-rise reinforced concrete in Vienna and Berlin, unreinforced masonry in Athens, Istanbul, and Naples. The conditioning period follows from the intensity measure of the matching ESRM20 fragility function. Surface amplification, spectral ordinates, and the damage and loss values reproduce Eqs.~\ref{eq:fragility} and~\ref{eq:aal} at the corresponding intensities, while the loss ordering reproduces the ESRM20 administrative-level reference. Batch cost is dominated by the per-site hazard integrals quantified below, with the downstream fragility, loss, and record-selection steps adding only tens of seconds per city, about 42~s per site for record selection. Per-city output is collected in Table~S3. Figure~\ref{fig:journeys} traces the resulting pair of journeys, namely the layer-by-layer passage of the Istanbul hazard query and the Naples masonry-loss query from this batch.

A first hazard computation at a fresh site runs between 37~s and 488~s on a single CPU core. At the fast end are low-complexity cratonic sites such as Dublin, while the slow end reflects the most source-dense Aegean sites such as Heraklion. Across the seventy-three-city benchmark, the median is 80~s. Once a site has been computed, repeat calls hit the session cache and return in under one second, whether the request is for a different building taxonomy, a different return period, or a follow-up retrofit analysis. In a typical interactive engineering session, then, only the first call carries the full probabilistic seismic hazard analysis cost. Subsequent record selection, fragility lookup, and sensitivity sweeps reuse the cached hazard curves.

\begin{figure}[htbp]
  \centering
  \includegraphics[width=\linewidth]{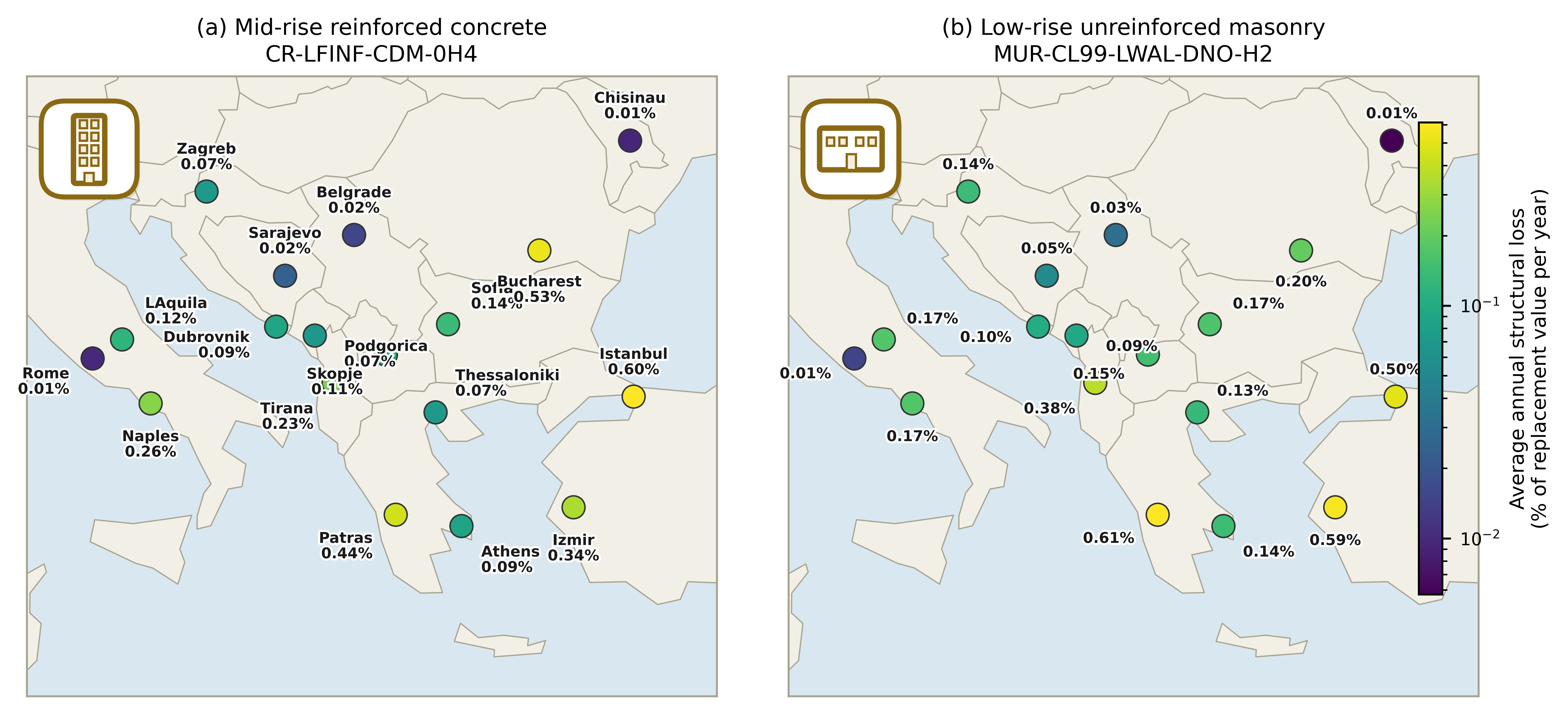}
  \caption{Average annual structural loss ratio at a south-eastern-European city panel for two representative residential classes: (a) a mid-rise reinforced-concrete frame (CR-LFINF-CDM-0H4) and (b) a low-rise unreinforced-masonry building (MUR-CL99-LWAL-DNO-H2), each computed by convolving the city's surface hazard curve with the matching ESRM20 vulnerability function at its fragility intensity measure (Eq.~\ref{eq:aal}; SA(1.0~s) and SA(0.3~s), respectively). The two panels share a single colour scale. The percentage is the expected annual structural repair cost as a fraction of replacement value. Each value is the average annual loss of a single representative building of the given class, evaluated at the city's own hazard and site conditions, not an aggregate over a portfolio; holding the building class fixed means the variation across cities reflects only hazard and site response. This single-building, single-site resolution is finer than the losses published with ESRM20, which are aggregated over the mixed building portfolio of each administrative unit.}
  \label{fig:loss}
\end{figure}

Risk replication scales to a regional panel. We drove the loss endpoint across a panel spanning Italy, the Balkans, Greece, Turkey, Romania, and Moldova for a fixed mid-rise reinforced-concrete class (CR-LFINF-CDM-0H4), which produces panel~(a) of Figure~\ref{fig:loss}. Re-running it for a low-rise unreinforced-masonry class (MUR-CL99-LWAL-DNO-H2), with the cached hazard reused at each site, produces panel~(b). For the reinforced-concrete class the structural AAL ratio ranges from 0.010\,\% to 0.60\,\% of replacement value per year, tracking the three-order-of-magnitude span in hazard. Because each class is evaluated at its own ESRM20 fragility intensity measure, namely SA(1.0~s) for the frame and SA(0.3~s) for the masonry, the two panels rank the cities differently. The comparison therefore exposes the spectral-shape sensitivity of risk rather than a uniform vulnerability offset. Every value in Figure~\ref{fig:loss} is the average annual loss of a single representative building of the chosen class, computed at the city's own hazard and site conditions; it is not an aggregate over a building portfolio. Holding the building class fixed across the panel ensures that the city-to-city variation comes only from hazard and site response, not from differences in the building stock. This single-building, single-site resolution is finer than the products published with ESRM20, which report losses aggregated over the mixed building portfolio of each administrative unit. The per-building, per-site loss is one of the capabilities not available from the EFEHR services (bold in Figure~\ref{fig:endpoints}).

\section{Discussion}

\subsection{Why hallucination is contained by design}

Hallucination, confident output that is nonetheless wrong, is the dominant failure mode of a language model in any technical setting. Recent surveys converge on a single mitigation: confine the model to planning and translation, and hand every numerical step to validated code \citep{lin2025,liY2025}. Our architecture follows this principle. The agent never produces a hazard curve, a fragility value, a damage probability, or a loss number. Because of this, hallucination in the seismological content of a response is prevented by design.

Two residual failure modes remain. In the first, the model could invent the arguments to a call, such as a fabricated city or an unphysical shear-wave velocity. Here the per-tool JSON Schema rejects the call before the handler runs \citep{hasan2026}. In the second, the model could skip a tool or misread its output. That mode is mitigated by the small catalogue, together with a shipped skill description that names the canonical call sequence for each user intent. A dedicated evaluation supports this account (Supplementary~S7). When the skill description was supplied to a locally served 7B model, its tool-selection accuracy rose from 86\,\% to 93\,\%, and the model stopped calling tools for requests the server cannot answer. On the more capable Claude Sonnet 4.6, accuracy reached 96\,\%, again with no calls to tools for unanswerable or vague requests.

Every response also carries the source-model branches, GMM labels, and data sources, with any approximation flag recorded in named provenance fields \citep{errico2025}. Because the record-selection workflow returns its actual filter bounds and error norms, departures from recommended ranges stay visible. The ground truth lives in the validated code and the EFEHR data, and the agent guides the user through it.

\subsection{Limitations and extensions}

The default branch subsets evaluate the dominant axes of the ESHM20 logic tree, but not every authorized branch. Where long-tail uncertainty matters, a richer subset can be requested through optional arguments. Two efficiency trade-offs operate within each branch and are folded into the replication residual: the area sources are discretized on a 20~km grid, and each tectonic region is integrated only out to a finite source-to-site distance cutoff (from 50~km for volcanic zones to 500~km for the deep Vrancea zone). Hazard is evaluated deterministically on the highest-weighted source and ground-motion branches, rather than by Monte-Carlo sampling of the full logic tree. This approximation is not an implementation defect; it is the principal source of the residual against the full-tree reference. The ground-motion override is not restricted to one model per tectonic region. Through the browser interface a user can supply a complete custom ground-motion logic tree with its own branches and weights. A few taxonomy strings lack a tabulated ESRM20 vulnerability function and fall back to a fragility-only path with a provenance flag. Finally, the topographic-slope shear-wave-velocity proxy \citep{heath2020} is uncertain at any single site, so a measured value should override it where available. The first-call hazard timings reported in the Results (37~s to 488~s per site, median 80~s) are for a single CPU core. Because OpenQuake supports distributed execution, these costs can be reduced substantially by running the engine in parallel on a high-performance-computing cluster, for example under a SLURM scheduler, which offers a practical route to interactive response times at the most source-dense sites. Denser sampling helps only when it is representative. Enlarging the source subset from three to five area-source branches raised the error at every site tested, because the twenty-one-branch area-source tree has no dominant member, so source-side convergence needs Monte-Carlo sampling at native weights rather than a larger top-weighted subset.

The layered architecture exposes several extension points. Within the hazard kernel, an event-based engine can be admitted for non-Poissonian time-dependent seismicity, as can a machine-learning surrogate \citep[such as][]{lehmann2025} for fast emulation. Because the protocol adapter is independent of both engine and domain logic, a successor protocol would primarily affect the topmost layer. The same layered organization, engineering-domain idiom, and provenance discipline are meant to be reusable for other regional hazard models, and in principle they extend to multi-hazard pipelines and real-time services, subject to the model- and domain-specific adaptations. Multi-agent extensions follow the same pattern. Here, separate agents encapsulate competing source models, GMM trees, or site-amplification assumptions, and a coordinator reconciles them; its inter-agent dispersion operationalizes epistemic uncertainty at the orchestration layer \citep{chenBao2025,liu2026,pantiukhin2025}.

A second perspective concerns the role assigned to the language model itself. Here the agent acts mainly as a coordinator. It parses user intent, plans a sequence of validated calls, reconciles their outputs, and renders the structured result, while numerical responsibility stays with the validated hazard code. Richer agentic roles fit the same layered design. A reinforcement-learning policy or a learned surrogate, trained on a library of full-logic-tree runs, could estimate the full-tree hazard without running every branch through the engine. One route is to learn which small subset of source-model and GMM branches suffices to reproduce the full-tree result at a given site. A second is to emulate the aggregated hazard curve directly and call OpenQuake only where the surrogate is uncertain. Similar surrogates could replace expensive subcomponents of disaggregation and scenario simulation. Self-improving systems, possibly multi-agent, could iterate over their own record selections, site-condition assumptions, or vulnerability mappings in response to validation feedback. All of this holds only if every learned component continues to expose validated, provenance-bearing inputs and outputs.

\section{Conclusions}

We have presented the first agentic interface for end-to-end probabilistic seismic hazard and risk analysis. It is an open-source server, addressable through the Model Context Protocol, that wraps the OpenQuake engine and the published European reference models behind twenty-four typed endpoints. Three usage modes are supported: replication, substitution, and extension. Benchmarked against the ESHM20 pan-European reference, the server reproduces 475-year spectral accelerations within a median of about 5\,\%, and the few high-residual sites trace to long-tail uncertainty from intermediate-depth seismicity rather than to implementation defects. A per-tectonic-region override interface accepts user-supplied analytic or machine-learning GMMs against the published regional tree at hazard-map level, providing a first-step comparison channel that complements traditional residual-based metrics. Conditional-spectrum record selection from the ESM database then closes the loop to time-history input. Because the layered architecture admits substitution of the protocol, the engine, the regional dataset, and the GMM logic tree independently, the design re-instantiates for any region, code framework, or open hazard engine. The result is both a working tool for the European earthquake engineering community and a reusable template for agentic interfaces to validated scientific computing.

\section*{Data and Resources}
\addcontentsline{toc}{section}{Data and Resources}

Source code, figure-generation scripts, and per-city verification outputs will be released with a Zenodo digital object identifier on acceptance. ESHM20 inputs are taken from the EFEHR GitLab project \texttt{eshm20} (\url{https://gitlab.seismo.ethz.ch/efehr/eshm20}), version v12e, with the computational bundle archived at \url{https://doi.org/10.12686/eshm20-oq-input} and the model overview at \url{https://doi.org/10.12686/a15}. ESRM20 inputs are taken from the \texttt{esrm20} repository, with the technical report at \url{https://doi.org/10.7414/EUC-EFEHR-TR002-ESRM20}; exposure data are taken from \texttt{esrm20\_exposure} and from \url{https://doi.org/10.5281/zenodo.4062044}; vulnerability and fragility tables from \texttt{esrm20\_vulnerability}. The OpenQuake engine is at \url{https://github.com/gem/oq-engine} \citep{pagani2014}. The USGS Global $V_{S30}$ Mosaic is at \url{https://earthquake.usgs.gov/data/vs30} \citep{heath2020}. The ESM database is at \url{https://esm-db.eu} \citep{mascandola2026}. Eurocode 8 site classes follow EN 1998-1. The ESHM20 reference spectra are from the EFEHR hazard service \citep{danciu2024}. The Model Context Protocol specification is at \url{https://modelcontextprotocol.io} \citep{anthropic2024}. Geocoding uses the Nominatim service at \url{https://nominatim.openstreetmap.org} with an offline fallback. All resources were last accessed in May 2026.

\section*{Code Availability}
\addcontentsline{toc}{section}{Code Availability}

The source code for the agentic seismic hazard and risk server, including the JSON-RPC handlers, the engineering-domain translator, the spatial cache, the figure-generation scripts, and the Streamlit interface, is openly available at \texttt{<GitHub URL --- will be provided upon acceptance>}.

\section*{Acknowledgments}
\addcontentsline{toc}{section}{Acknowledgments}

We thank the EFEHR consortium and the GEM Foundation for openly releasing ESHM20, ESRM20, and OpenQuake hazardlib, and the maintainers of the ESM database waveforms. This research was carried out in the MINERVE project number DOS0186108. The MINERVE project is supported by the French government within the France 2030 framework: CORIFER AAP1 -- Projet ``MINERVE''. The authors thankfully acknowledge this support.

\section*{CRediT authorship contribution statement}
\addcontentsline{toc}{section}{CRediT authorship contribution statement}

\textbf{Sreenath Vemula:} Writing -- original draft, Visualization, Validation, Software, Methodology, Investigation, Formal analysis, Data curation, Conceptualization. \textbf{Pierre Jehel:} Resources, Project administration, Funding acquisition. \textbf{Fabrice Cotton:} Writing -- review \& editing, Supervision. \textbf{Filippo Gatti:} Writing -- review \& editing, Resources, Supervision, Project administration.

\section*{Declarations}
\addcontentsline{toc}{section}{Declarations}

\noindent\textbf{Conflict of interest.}\quad The authors declare no competing interests.

\medskip
\noindent\textbf{Declaration of generative AI in scientific writing.}\quad While preparing the manuscript, the authors used generative AI tools such as Claude Sonnet 4.6 and Opus 4.8 to improve the readability and correct the language. After using these tools, the authors reviewed and edited the content as required and take full responsibility for the content of the published article.

% ---------- References ----------
\nocite{*}
\bibliography{references}

\end{document}